\documentclass{amsart}

\title[The standard model and gravity in $E_8$]{On possible embeddings of the standard models of particle physics and gravity in $E_8$}
\author{Robert Arnott Wilson}
\date{First draft: 20th April 2024; this version; 6th May 2024.}

\address{Queen Mary University of London}
\email{r.a.wilson@qmul.ac.uk}

\newcommand{\RR}{\mathbb R}
\newcommand{\CC}{\mathbb C}
\newcommand{\HH}{\mathbb H}
\newcommand{\lie}{\mathfrak}
\newcommand{\rep}{\mathbf}

\begin{document}
\begin{abstract}
I investigate the structure of $E_8$ under the action of the subalgebra/subgroup $A_1+G_2+C_3$,
as a potential route to unification of the fundamental forces of nature into a single algebraic structure.
The particular real form $E_{8(-24)}$ supports a decomposition into compact $G_2$ plus split $A_1+C_3$,
which allows a restriction from $G_2$ to $SU(3)$ for the strong force, 
together with split $SL_2(\RR)$ to break the
symmetry of the weak interaction and give mass to the intermediate vector bosons. The factor $C_3$ contains various copies of the
Lorentz group $SL_2(\CC)$ and extends the `spacetime' symmetries to the full group of symplectic symmetries of 
real $3+3$-dimensional phase space. 

Restricting $G_2$ to the Standard Model $SU(3)$ extends $C_3$ to $A_5$, in the real form $SU(3,3)$, 
acting on a complex phase space that includes both momentum and current.
There is then a natural restriction from $SU(3,3)$ to $SO(3,3)$, describing the action of $SL_4(\RR)$ on phase space.
The resulting action of $SL_4(\RR)$ on $E_8$ includes tensors that are equivalent to the stress-energy tensor,
the Ricci tensor and the Riemann tensor, and therefore permits the formalism of general relativity to be developed
inside $E_{8(-24)}$.  The model then suggests unexpected and perhaps subtle ways in which general relativity and particle physics may 
be forced to modify each other, in order to produce a unified theory. 
\end{abstract}
\maketitle

\section{Introduction}
A number of $E_8$ models of fundamental physics have been proposed in recent years \cite{MDW,Lisi,Lisi2,Chester,chirality}, but none of them has been sufficiently compelling
to persuade large numbers of people that they are useful. The key issue is how to split up the symmetries of $E_8$ to get something that looks like the Standard Model,
and in particular, how to do this in a reasonably `natural' way. In addition there are a number of technical issues which cause a lot of trouble, particularly to do with
complex structures and chirality, and with implementing three generations of fermions when there appears on the face of it to be only enough room for two \cite{DG}.

The approach to `naturality' taken in \cite{MDW} is to take the Freudenthal--Tits `magic square' \cite{Freud,Tits,BS} as a guide. The Lie structure of 
the magic square is
\begin{align}
\begin{array}{cccc}
A_1 & A_2 & C_3 & F_4\cr
A_2 & A_2+A_2 & A_5 & E_6\cr
C_3 & A_5 & D_6 & E_7\cr
F_4 & E_6 & E_7 & E_8
\end{array}
\end{align}
and most emphasis has been put on the fourth row and the fourth column, where the exceptional Lie groups are found. However, the other entries in the table are also interesting.
For example, the route taken in \cite{MDW} from top left to bottom right goes via $A_2$, $A_5$ and $E_6$ in the second row (see also \cite{MD,Todorov}).

The corresponding `magic square' for $2\times 2$ matrices is
\begin{align}
\begin{array}{cccc}
B_0 & B_1 & B_2 & B_4\cr
B_1 & D_2 & D_3 & D_5\cr
B_2 & D_3 & D_4 & D_6\cr
B_4 & D_5 & D_6 & D_8  
\end{array}
\end{align}
where I 
use the notation $B$ and $D$ to emphasise that these are all spin groups, with $B_n=Spin(2n+1)$ and $D_n=Spin(2n)$.
However, I would like to
draw particular attention to the isomorphisms between spin groups (types $B$ and $D$) and unitary (type $A$) and symplectic (type $C$) groups:
\begin{align}
B_1 & =A_1, \cr D_2 & = A_1+A_1,\cr B_2 & = C_2,\cr D_3 & =A_3.
\end{align}
All of the exceptional isomorphisms listed here arise from the triality automorphism of $D_4$, which links
together all the groups in the table that are not in the fourth row or fourth column. 

The isomorphism $B_1=A_1$ gives the basic fact that $Spin(3) \cong SU(2)$, which is the foundation of quantum mechanics (QM),
while $D_2=A_1+A_1$ extends this
to $Spin(3,1)\cong SL(2,\CC)$ for relativistic QM.  
The isomorphism $D_3=A_3$ 
is needed to explain the complex structure of the Dirac algebra, as the complex Clifford algebra of Minkowski spacetime,
since the smallest real Lie algebra that contains all the Dirac matrices, including $\gamma_5$, is of this type.
This isomorphism also
plays a significant role in various Grand Unified Theories (GUTs)
from 1974 onwards \cite{PatiSalam,twistors,twistorlectures}, in various real forms
including $Spin(6)\cong SU(4)$, $Spin(4,2)\cong SU(2,2)$ and $Spin(3,3)\cong SL_4(\RR)$.
The isomorphism $B_2=C_2$
similarly lies at the heart of the AdS/CFT correspondence \cite{Maldecena},
and describes the Lie algebra of the even part of the Dirac algebra. 

Extending then to the $3\times 3$ case, we extend $C_2$ to $C_3$, which looks a lot more `natural' than
extending $B_2$ to $C_3$. In this note, therefore, I concentrate on the third row of the magic square \cite{DMWE7a,DMWE7},
and especially the first group, $Sp_6(\RR)$, that has a classical interpretation as the symmetry group of phase space \cite{deGosson}.
The second group extends this to a complex phase space, so that both momentum and current
can be included, as well as the position of both mass and charge. This extends the symplectic group to $Sp_6(\CC)\cong SU(3,3)$,
whose centralizer in $E_8$ is $SU(3)\times SL_2(\RR)$. The latter is a real form of the gauge group 
$SU(3)\times SU(2)$ of the strong and weak nuclear forces, and I shall argue in this paper 
that the group $SU(3,3)$ describes both classical and quantum
forms of both electromagnetism and gravity.

\section{Embedding $C_2$ in $D_8$}
\label{embedC2}
In the semi-split version of the magic square, both the compact and split real forms of $C_2$ occur, but because I want to use $C_2$ to implement the Dirac algebra,
I want the split form, that is the group $Spin(3,2)\cong Sp_4(\RR)$. Embedding into $Spin(12,4)$ we see the centralizer $Spin(9,2)$, which we can split into
three pieces, $Spin(6)$, $Spin(3)$ and $Spin(2)$, if we want to get the Standard Model gauge groups $SU(3)$, $SU(2)$ and $U(1)$. However, the reason for this splitting
will not become clear until we consider the embedding of $C_3$ in $E_8$. 

In the notation of \cite{MDW,WDM}, we need to choose a copy of the split quaternions $\HH'$ in the split octonions, say the copy with basis $U,K,L,KL$.
Then the corresponding copy of $Spin(3,2)$ acts on the indices $u:=1,U,K,L,KL$, leaving $IL, JL$ for $Spin(2)$, and $i,j,k,il,jl,kl$ for $Spin(6)$ and/or $SU(3)$,
so that $Spin(3)$ acts on $l,I,J$. In particular we see some symmetry-breaking for $Spin(3)$, 
generated by $X_{lI}$, $X_{lJ}$ and $D_{I,J}$, already in the notation. Now for 
the Dirac part of the algebra, the labels $u,U,K,L,KL$ correspond to the matrices $\gamma_1,\gamma_2,\gamma_3,\gamma_0$ and $\gamma_5$ respectively, and the
products of pairs of these gamma matrices generate the algebra $\lie{so}(3,2)$: 
\begin{align}
&X_1, X_K, D_K, \cr &X_L, D_L, D_{K,L}, \cr &X_{KL}, D_{KL}, D_{K,KL}, D_{L,KL},
\end{align}
in which the first row contains the rotations in $SL_2(\CC)$, and the second row contains the boosts, while the third row contains a Lorentzian $4$-vector.

Since this algebra is quaternionic rather than octonionic, its elements can be written as ordinary $2\times2$ anti-Hermitian
quaternion matrices, which make it easier to understand the structure.
The $X$s are off-diagonal, the single-index $D$s are diagonal traceless, and the double-index $D$s add the imaginary traces:
\begin{align}
&\begin{pmatrix}0&1\cr -1&0\end{pmatrix},
\begin{pmatrix}0&K\cr K&0\end{pmatrix},
\begin{pmatrix}K&0\cr 0&-K\end{pmatrix},\cr
&\begin{pmatrix}0&L\cr L&0\end{pmatrix},
\begin{pmatrix}L&0\cr 0&-L\end{pmatrix},
\begin{pmatrix}KL&0\cr 0&KL\end{pmatrix},\cr
&
\begin{pmatrix}0&KL\cr KL&0\end{pmatrix},
\begin{pmatrix}KL&0\cr 0&-KL\end{pmatrix},
\begin{pmatrix}L&0\cr 0&L\end{pmatrix},
\begin{pmatrix}K&0\cr 0&K\end{pmatrix}
\end{align}
These matrices can be regarded as a split quaternionic version of the Pauli matrices, which occur in the first row, in the mathematicians' anti-Hermitian convention
rather than the physicists' Hermitian convention. The last matrix in the third row is the scalar matrix that is the product of the three Pauli matrices (Hermitian convention). 

\section{Extending to $C_3$}
\label{embedC3}
To extend from $C_2$ to $C_3$ we have to generalise the $X$ terms to $Y$ and $Z$, and the $D$ terms to $E$:
\begin{align}
Y_1, Y_K, Y_L, Y_{KL},\cr
Z_1, Z_K, Z_L, Z_{KL},\cr
E_K, E_L, E_{KL}.
\end{align}
Equivalently, we extend from $2\times 2$ matrices to $3\times 3$, to obtain a split quaternionic version of the Gell-Mann matrices \cite{GellMann}. 
It turns out that the `scalar' $2\times2$ matrices extend to traceless $3\times 3$ matrices, in the same way that the Gell-Mann matrices are traceless.
In the notation of \cite{WDM}, this condition is enforced by the identity:
\begin{align}
D_p+E_p+F_p=0
\end{align}
for all single index $p$, here $K$, $L$ and $KL$.
 I make no claims as to how these Gell-Mann matrices should be interpreted, or whether they have anything to do with
the Gell-Mann matrices used in QCD \cite{QCD}.

In fact, the compact part of this group $Sp_6(\RR)$ is a copy of $U(3)$ generated by all the elements that have an even number of copies of $L$ in their labels:
\begin{align}
X_1,X_K,Y_1,Y_K,Z_1,Z_K, D_K, E_K, D_{L,KL}
\end{align}
where again the double-index $D$s add an imaginary trace to the matrices. 
As matrices, these are just the anti-Hermitian matrices over the complex subalgebra $\CC$ of $\HH'$.
There are also two copies of $GL_3(\RR)$ obtained by replacing $K$ by $L$ or $KL$ 
(or indeed any linear combination of the two):
\begin{align}
&X_1,X_L,Y_1,Y_L,Z_1,Z_L, D_L, E_L, D_{K,KL};\cr
&X_1,X_{KL},Y_1,Y_{KL},Z_1,Z_{KL}, D_{KL}, E_{KL}, D_{K,L}.
\end{align}
These are anti-Hermitian matrices over the respective copies of the split complex numbers $\CC'$ in $\HH'$.
Leaving off the double-index $D$s restricts to $SL_3(\RR)$, and the subtle relationships between these two copies of $SL_3(\RR)$ and $SU(3)$
will play an important role in this paper.

\section{The centralizer of $C_3$}
\label{C3cent}
The centralizer of $Sp_6(\RR)$ in $E_8$ is a group of type $A_1+ G_2$, in which the copy of $G_2$ is compact, acting on the indices $i,j,k,l,il,jl,kl$, and the copy of
$A_1$ is split, acting on the indices $I,J,IL,JL$ in one of its chiral spinor (weak isospin?) representations. This copy of $A_1$ is obviously not the same as the copy, 
acting on $l,I,J$, that we suggested in Section~\ref{embedC2}
simply by looking in $D_8$. However, we effectively chose a copy of $\lie{su}(2)+\lie{u}(1)$ acting on $l,I,J,IL,JL$, so that by `mixing' $D_{I,J}$ with $D_{IL,JL}$ we
obtain one of the elements of the centralizer. To get the rest we need to replace $X_{lI}$ and $X_{lJ}$ by one combination of $D_{I,IL}$ and $D_{J,JL}$ and another
combination of $D_{I,JL}$ and $D_{J,IL}$. 

In other words, the conversion between these two copies of $A_1$, one of which is compact and the other split, is
very reminiscent of the `symmetry-breaking' of the weak $SU(2)$ in the Standard Model, that converts from a `primordial' massless $SU(2)$ to $SL_2(\RR)$ via the
complexification $SL_2(\CC)$, in order to give the intermediate vector bosons non-zero masses. Thus the $E_8$ model provides a fundamental mathematical reason for
this symmetry-breaking, namely enforcement of the condition that the gauge group must commute not only with $D_2$, as the Coleman--Mandula theorem \cite{Coleman}
requires, or
with $C_2$, as the embedding in $D_8$ requires, but with the whole of $C_3$.

To see the details of how these two copies of $A_1$ are related to each other, we can work in the group they generate, which is another copy of $Spin(3,2)$, acting on the
labels $l,I,J,IL,JL$, and commuting with the first copy. The ten dimensions of this group are represented by
\begin{align}
&X_{lI}, X_{lJ}, D_{I,J}, D_{IL,JL},\cr
& X_{lIL}, X_{lJL}, D_{I,IL}, D_{I,JL}, D_{J,IL}, D_{J,JL}.
\end{align}
Since we are genuinely using the octonions at this point, it is not possible to write these elements of the algebra as ordinary matrices.
However, the notation of \cite{WDM} may help to visualise what is going on.
The generators of the two copies of $A_1$ are related as follows (where the signs are determined by the chirality, or equivalently
by the embedding in $G_2$):
\begin{align}
\label{A1gens}
D_{I,J} & \rightarrow  D_{I,J}- D_{IL,JL}\cr
X_{lI} & \rightarrow D_{I,IL}- D_{J,JL}\cr
X_{lJ} & \rightarrow D_{J,IL}+ D_{I,JL}.
\end{align}

Turning now to the remaining factor, which is presumably related to the strong force, we compare the group $Spin(6)$ that appears in the centralizer of $Spin(3,2)$, with
the group $G_2$ that appears in the centralizer of $Sp_6(\RR)$. After replacing the original (unbroken symmetry) copy of $A_1$ by the chiral copy on $I,J,IL,JL$, we
no longer require the label $l$ for the weak interaction, which can be added to the strong force, to extend the gauge group $SU(3)$ acting  on $3+3$ colours and anti-colours
to $G_2$ acting on $7$ `colours'. This extension is reminiscent of the Pati--Salam model, which uses $SU(4)$ for four colours and four anti-colours, but is
group-theoretically completely different.

 Notice also that we have a chiral pair of left-handed and right-handed copies of $SL_2(\RR)$, so that there is a close parallel 
between the two models:
\begin{align}
SU(4) \times SU(2)_L \times SU(2)_R\cr
G_2 \times SL_2(\RR)_L \times SL_2(\RR)_R.
\end{align}
Generators for $SL_2(\RR)_R$ in this case are obtained by changing the signs in (\ref{A1gens}).
But this group $SL_2(\RR)_R$ is not contained in $Sp_6(\RR)$, so we do not find it useful,
and shall not use it.
In Section~\ref{A5extend} we consider a different type of `right-handed' $SL_2(\RR)$, whose
generators are given in (\ref{A1Rgens}), which is contained in $Sp_6(\RR)$, and has more useful
symmetry properties.

Here we have a total of $20$ degrees of freedom, compared to $21$ in the Pati--Salam model. 
Therefore our full model is of type $C_3+A_1+G_2$, based on the group
 \begin{align}
 Sp_6(\RR) \times SL_2(\RR) \times G_2,
 \end{align}
 with the first factor generalising the Lorentz group, the second factor representing a real form of weak $SU(2)$, and the third factor generalizing strong $SU(3)$.
 
 \section{Extending to $A_5$}
 \label{A5extend}
 The occurrence of $G_2$ in the decomposition $A_1+G_2+C_3$, rather than $A_2$, that we would expect for the strong force, suggests that we should move to the
 second group in the third row of the magic square, of type $A_5$, and the associated decomposition $A_1+A_2+A_5$ of $E_8$. The particular real forms that arise are
 \begin{align}
 SL_2(\RR) \times SU(3) \times SU(3,3),
 \end{align}
 which is analogous to the decomposition
 \begin{align}
 SU(2) \times SL_3(\RR) \times SL_3(\HH)
 \end{align}
 studied in \cite[Section II.C]{MDW}. I propose these different real forms as a potentially closer match to the Standard Model, since the compact group $SU(3)$
 is more suitable for massless gluons, while the split group $SL_2(\RR)$ has two boosts that are suitable for masses of the intermediate vector bosons. This is because mass is usually introduced by complexifying the compact gauge group, precisely in order to generate boosts.
 
 To extend $Sp_6(\RR)$ to $SU(3,3)$ we add $14$ dimensions, all involving the complex structure $l$. The analogous $2\times 2$ extension is from
 $Spin(3,2)$ acting on $u,U,K,L,KL$ to $Spin(4,2)$ acting on $l,u,U,K,L,KL$, so that the five new elements are
 \begin{align}
 X_l,D_l,X_{lK},X_{lL},X_{lKL}
 \end{align}
 Since we are not really using the octonions here, all these elements can be written as $2\times 2$ anti-Hermitian matrices over
\begin{align}
\CC \otimes \HH' = \langle u,l\rangle \otimes \langle U,K,L,KL\rangle
\end{align} 
so that the new matrices are
\begin{align}
\begin{pmatrix} 0&l\cr l&0 \end{pmatrix},
\begin{pmatrix} l&0\cr0&-l \end{pmatrix},
\begin{pmatrix} 0&lK\cr -lK&0 \end{pmatrix},
\begin{pmatrix} 0&lL\cr -lL&0 \end{pmatrix},
\begin{pmatrix} 0&lKL\cr -lKL&0 \end{pmatrix}.
\end{align}
 To get the whole of $SU(3,3)$, therefore, we need to extend from $2\times 2$ to $3\times 3$ matrices, 
 which means adding in the corresponding $Y$s and $Z$s, and one $E$:
 \begin{align}
 E_l,Y_l,Y_{lK},Y_{lL},Y_{lKL},\cr
 Z_l,Z_{lK},Z_{lL},Z_{lKL}.
\end{align}

 The `diagonal' part of this group is $U(1)\times U(1)\times SL_2(\RR)\times SL_2(\RR)\times SL_2(\RR)$, an $11$-dimensional group generated by 
 \begin{align}
 D_l,E_l, D_K,E_K,D_{L,KL},D_L,E_L, D_{K,KL},D_{KL},E_{KL}, D_{K,L},
 \end{align}
 and the off-diagonal part consists of $8$ dimensions each of elements of type $X$ (bosonic), $Y$ and $Z$ (fermionic). 
 Adding any one of these three types gives a group $U(1)\times SL_2(\RR)\times SU(2,2)$ of dimension $19$.
 The subgroup $Sp_6(\RR)$
 loses the first two diagonal elements of type $D$, and half of the off-diagonal elements of types $X,Y,Z$.
 In this case the $D$s and $X$s generate a group $SL_2(\RR)\otimes Sp_4(\RR)$ of dimension $13$. This last copy of $SL_2(\RR)$ is
 generated by
 \begin{align}
 E_K-F_K, E_L-F_L, E_{KL}-F_{KL},
 \end{align}
 or equivalently by
 \begin{align}
 \label{A1Rgens}
& D_{I,J}+D_{IL,JL}+D_{KL,L},\cr
& D_{I,IL}+D_{J,JL}+D_{K,KL},\cr
& D_{J,IL}+D_{JL,I}+D_{K,L}.
 \end{align}
 and is a type of `right-handed' $SL_2(\RR)$, as described in Section~\ref{C3cent}.
 
 The particular real form $SU(3,3)$ suggested here as a (huge) generalisation of the Lorentz group, from $6$ dimensions to $35$, is closely related to Penrose twistors,
 since the corresponding entry in the magic square of $2\times 2$ matrix groups is $SU(2,2)$. 
 In other words, $SU(3,3)$ combines the group $Sp_6(\RR)$ of symmetries of phase space with the group $SU(2,2)$ of symmetries of twistors, into a single
 symmetry group. If this mathematical unification can lead to a physical unification, then it could have far-reaching consequences for the fundamental theory.
 
 It should be noted that $SU(2,2)$ embeds in $Spin(12,4)$ in two different ways, as $Spin(2,4)$ and as $Spin(4,2)$, centralizing $Spin(10)$ and
 $Spin(8,2)$ respectively. The former was used in \cite{chirality} and extended
 to $SU(2,3)\times SU(5)$ in an attempt to understand how twistors relate to $E_8$ models. Here we use the latter instead, so that the centralizer 
 splits
 as $Spin(6)\otimes Spin(2,2)$ to give a different real form of the Georgi $Spin(10)$ GUT, and a different embedding of the twistors into $E_8$.
 Comparing with \cite{MDW}, we see that the latter uses $Spin(3,3)\otimes Spin(4)$ inside $Spin(7,3)$,
 as yet another real form of $Spin(10)$. 
 It is worthwhile considering which
 real form is most appropriate. 
 In the Standard Model, the strong force $SU(3)$ is definitely compact, and the mediators are correspondingly massless, but the
 weak force $SU(2)$ is definitely not compact, because the complexification is used to allow the mediators to be massive. So of the three choices $Spin(10)$,
 $Spin(7,3)$ and $Spin(8,2)$, only the last has a reasonable chance of agreeing with the Standard Model.
 
 The group $Sp_6(\RR)$ is studied in detail in \cite{CAMPS}, embedded in a different real form of $SU(6)$, namely $SL_3(\HH')$. It is straightforward to translate
 that work into $SU(3,3)$, simply by multiplying the Hermitian matrices by $l$ to make them anti-Hermitian over $\CC\otimes \HH'$. 
 However, any interpretation in \cite{CAMPS} that is based on the particular real form is suspect. The study of $SL_3(\HH)$ in \cite{MDW}
 is also not difficult to translate into, or out of, these other two real forms. But again, the interpretation offered in \cite{MDW}
 is quite different from the interpretation I offer here.
 The great advantage of using $SU(3,3)$ rather than $SL_3(\HH')$ 
 is that it removes the contradiction with general relativity
 that was apparent in \cite{CAMPS} (see Section~\ref{gravity} below).

 \section{Representations}
 Let us first look at the restriction  to $A_1+A_2+A_5$ of the adjoint representation of $E_8$. The real constituents for the real form $SL_2(\RR)\times SU(3)\times SU(3,3)$
 are as follows:
 \begin{align}
 \rep3 & = \rep3\otimes \rep1\otimes \rep1\cr
 \rep8 & = \rep1\otimes \rep8 \otimes \rep1\cr
 \rep{35} & = \rep1\otimes \rep1\otimes \rep{35}\cr
 \rep{40} & = \rep2\otimes \rep1\otimes \rep{20}\cr
 \rep{90} & = \rep1 \otimes \rep3_\CC\otimes_\CC \rep{15}_\CC\cr
 \rep{72} & = \rep2\otimes \rep3_\CC\otimes_\CC \rep6_\CC
 \end{align}
 The first three constituents are the adjoint representations of the three factors, and the last three involve the real weak doublet representation $\rep2$ and the 
 complex colour triplet representation $\rep3_\CC$. The representations of $SU(3,3)$ are the natural $6$-dimensional complex representation $\rep6_\CC$, its
 anti-symmetric square $\rep{15}_\CC$
 and its anti-symmetric cube (real $\rep{20}$). 
 
 Restricting to $SU(2,2)$ to separate bosonic and fermionic representations we have
 \begin{align}
\rep 6_\CC & \rightarrow (1+1) + 4\cr
 \rep{20} & \rightarrow (6+6)+(4+4)\cr
 \rep{15}_\CC & \rightarrow (1+6)+(4+4)
 \end{align}
 which gives us a total of $16+48+48 =112$ dimensions of spinors, of which $48$ are right-handed and $64$ are left-handed.
 The left-handed spinors split $16+48$ into leptons and quarks, while the right-handed spinors here represent only quarks.
 The remaining $16$ dimensions of right-handed spinors lie inside the group $SU(3,3)$. In order to allocate some of these spinors to right-handed electrons,
 therefore, we need to break the symmetry back down to $Sp_6(\RR)$. This gives us $8$ dimensions of right-handed lepton spinors,
 compared to $16$ for left-handed leptons, which is the ratio that we expect.
   
 One way to study the splitting of $E_8$ into representations of $A_1+C_3+G_2$ is to embed it first in $F_4+ G_2$, where we have a decomposition
 \begin{align}
 \rep{248} = \rep{52} + \rep{14} + \rep{26}\otimes \rep7.
 \end{align}
 Then we restrict from $F_4$ to $A_1+C_3$ to get
 \begin{align}
 \rep{52} & = \rep3+\rep{21}+\rep2\otimes \rep{14}b\cr
 \rep{26} & = \rep2\otimes \rep6 + \rep{14}a
 \end{align}
 where $\rep6$ is the natural representation of $Sp_6(\RR)$, and the other representations are defined by
 \begin{align}
  \Lambda^2(\rep6) & =\rep1+\rep{14}a,\cr
  S^2(\rep6) & = \rep{21},\cr
  \Lambda^3(\rep6) & =\rep6+\rep{14}b.
  \end{align}
 Here, $\Lambda^2(\rep6)$ has a natural structure as a Jordan algebra, while $S^2(\rep6)$ has a natural structure as a Lie algebra.
 An alternative way to see these splittings is by restriction from $SU(3,3)$:
 \begin{align}
 \rep6_\CC & \rightarrow \rep6\cr
 \rep{15}_\CC & \rightarrow \rep1 + \rep{14}a\cr
 \rep{20} & \rightarrow \rep6 + \rep{14}b\cr
 \rep{35} & \rightarrow \rep{21} + \rep{14}a
 \end{align}
 
 As representations of $A_1+G_2+C_3$ we have the following irreducible constituents of $E_8$:
 \begin{align}
\rep3 &=  \rep3\otimes \rep1\otimes \rep1\cr
 \rep{14} &= \rep1\otimes \rep{14}\otimes \rep1\cr
\rep{21} &= \rep1\otimes \rep1\otimes \rep{21}\cr
 \rep{28} & = \rep2\otimes \rep1 \otimes \rep{14}b\cr
\rep{98} & = \rep1\otimes \rep7\otimes \rep{14}a\cr
\rep{84} & = \rep2\otimes \rep7\otimes \rep6
 \end{align}
 All the right-handed spinors (including the electrons) are now inside the $\rep{98}$, while the left-handed lepton spinors have been split $8+8$
 between the $\rep{28}$ and the $\rep{84}$,
 as a result of the breaking of $\rep{20}$ into $\rep{14}b+\rep6$. This curious phenomenon will no doubt repay closer scrutiny. It looks at first sight like
 a distinction between (massless) neutrinos and (massive) electrons, but that is not consistent with the identification of the $A_1$ factor as
 acting on weak doublets. Hence one or other of these suggested interpretations has to change.
 
 This analysis gives us a total of five different types of spinors:
 \begin{itemize}
 \item $8$ dof in $28$, left-handed leptons;
 \item $8+48$ dof in $98$, right-handed leptons and quarks;
 \item $8+48$ dof in $84$, left-handed leptons and quarks.
 \end{itemize}
 In total, then, there are $24$ dof for leptons, or $6$ Weyl spinors, compared to the $9$ that are usually expected for three generations. Similarly,
 there are $96$ dof for quarks, or $24$ Weyl spinors, compared to the $36$ that are usually expected. Thus we need a mechanism similar to that
 proposed in \cite{MDW} for reducing the number of independent spinors required by one-third. This is not surprising, of course, as it is well-known that the
 standard interpretation requires $180$ dof for spinors \cite{DG}. However, a \emph{discrete} symmetry of order $3$ can be implemented in a $2$-dimensional
 real space using the symmetries of an equilateral triangle, so there is no theoretical reason why a $3$-space is needed for this symmetry.
 
 An alternative interpretation is that this model describes only the first-generation fermions, since we have broken the generation symmetry
 by choosing $K$. In that case, we have not too few spinors, but too many. In particular, half of the spinors that we assumed were leptons are
 actually associated with quarks under the action of $G_2$. Hence it may be more reasonable to assume that they are (first generation)
 baryons, i.e. proton and neutron. This assumption allows us to use the representation $\rep{20}$ to describe all `normal' matter, made out
 of electrons, protons and neutrons, while leaving more exotic particles to be described by other representations.

 \section{Restricting to $A_2+A_2$}
 An alternative strategy for producing spinors for right-handed electrons is to restrict from $A_5$ to $A_2\times A_2$ instead of $C_3$. This extends the centralizer from
 $A_1+A_2$ to $A_2+A_2$, and gives rise to the following subgroup of $E_{8(-24)}$:
 \begin{align}
 SL_3(\CC) \times SU(3) \times SL_3(\RR).
 \end{align}
 This group provides an obvious embedding of the Lorentz group $SL_2(\CC)$ in $SL_3(\CC)$, defining the splitting into fermions and bosons, and offers various
 possibilities for $SL_2(\RR)$ or $SO(3)$ for the weak force. However, $SL_3(\RR)$ acts identically on vectors and both types of spinors, so does not provide
 any obvious way to implement the chirality of the weak force. We considered this possibility in the work that led to \cite{MDW}, but did not find a way to
 make it work. Of course, that does not necessarily mean that it cannot be done.
 
 The group $SL_3(\CC)$ is generated by the $16$ elements
 \begin{align}
 D_l, D_L,X_1,X_l,X_L,X_{lL},\cr
 E_l,E_L, Y_1,Y_l,Y_L,Y_{lL},\cr
 Z_1,Z_l,Z_{L},Z_{lL}
 \end{align}
 in which the top row is the Lorentz group $SL_2(\CC)$, centralized by a complex scalar generated by
 \begin{align}
 E_L-F_L & =D_L+2E_L,\cr
 E_l-F_l & =D_l+2E_l.
 \end{align}
 The group $SU(3)$ acts on the labels $i,j,k,il,jl,kl$, identically on $X$, $Y$ and $Z$. The group $SL_3(\RR)$ acts similarly on the labels $I,J,K,IL,JL,KL$.
 The $120$ spinors therefore split as $24+24+72$, in which the $72$ have both an $i,j,k$ and an $I,J,K$ in the label, and the $24$s have one or the other.

 We now have to break the symmetry of $I,J,K$ in order to separate left-handed and right-handed spinors, so let us separate $I,J$ from $K$ to break
 $72=24+48$ and one of the $24=8+16$. This gives us a splitting of quarks in $24+24+48$, such that $24+24$ are right-handed, and $48$ are left-handed.
 Similarly, the leptons split as $8$ right-handed and $16$ left-handed, again in agreement with the Standard Model. 
 To be more explicit, we give the labels in the form of a table, with the labels for the $C_3$ version for comparison:
 \begin{align}
 \begin{array}{r|cc|cc|}
 \mbox{RH l} & u,l & K,KL& u,l & U\pm KL,K\pm L\cr
 \mbox{LH l} & u,l & I,J,IL,JL & u,l & I,J,IL,JL\cr
 \mbox{RH q} & i,j,k,il,jl,kl & U,K,L,KL & i,j,k,il,jl,kl & U,K,L,KL\cr
 \mbox{LH q} & i,j,k,il,jl,kl & I,J,IL,JL & i,j,k,il,jl,kl & I,J,IL,JL 
 \end{array}
 \end{align}
 The actual signs for the right-handed leptons in the $C_3$ case are different in the $Y$ and $Z$ spinors, but only one sign occurs in each case.
 The allocation of individual particles in the $C_3$ and $A_2+A_2$ cases is not necessarily the same, but the overall picture is very similar.
 But only the $C_3$ case has the projection with $U-KL$ that corresponds to $1-\gamma_5$ in the Standard Model. This appears to be
 a decisive vote in favour of the $C_3$ model over the $A_2+A_2$ model, at least 
 for the Dirac algebra. 
 
 On the other hand, the Dirac equation is probably better modelled with $A_2+A_2$, where there are $12$ Lorentzian $4$-vectors available for the
 gamma matrices. Nine of these extend $Spin(3,1)$ to $Spin(4,1)$, and therefore have an obvious place to put a mass that is defined by the
 Einstein mass equation:
 \begin{align}
 \begin{array}{ccc}
D_I,X_I,X_{lI},D_{I,L},\cr
D_J,X_J,X_{lJ},D_{J,L},\cr
D_K,X_K,X_{lK},D_{K,L}
 \end{array}
 \begin{array}{ccc}
 X_i,D_i,D_{i,l},X_{iL}\cr
 X_j,D_j,D_{j,l},X_{jL},\cr
 X_k,D_k,D_{k,l},X_{kL}
 \end{array}
 \begin{array}{ccc}
 X_{il},D_{il},D_{il,l},X_{ilL}\cr
 X_{jl},D_{jl},D_{jl,l},X_{jlL},\cr
 X_{kl},D_{kl},D_{kl,l},X_{klL}
 \end{array}
 \end{align}
 The first column look like electrons, and the others look like quarks, although it is also possible to take arbitrary linear combinations.
 The other three extend $Spin(3,1)$ to $Spin(3,2)$, and therefore do not have an Einstein/Dirac mass term:
 \begin{align}
 \begin{array}{ccc}
D_{IL},X_{IL},X_{lIL},D_{IL,L},\cr
D_{JL},X_{JL},X_{lJL},D_{JL,L},\cr
D_{KL},X_{KL},X_{lKL},D_{KL,L}
 \end{array}
 \end{align}
 These are presumably neutrinos.
 \section{Symmetry-breaking}
 Nevertheless, there are many questions remaining about the differences between $C_3$ and $A_2+A_2$, particularly concerning the physical interpretations,
 and the reason for the symmetry-breaking from $SL_3(\RR)$ to $SL_2(\RR)$. In order to bring the questions into focus, it is useful to consider
 the square of groups $A_2$, $A_2+A_2$, $C_3$ and $A_5$, together with the corresponding block of the $2\times 2$ magic square, and the centralizers.
 \begin{align}
 \begin{array}{|cc|}
 \hline
 SL_3(\RR) & SL_3(\CC)\cr
 Sp_6(\RR) & SU(3,3)\cr\hline
 \end{array} \qquad
 \begin{array}{|cc|}
 \hline
 Spin(2,1) & Spin(3,1)\cr
 Spin(3,2) & Spin(4,2)\cr\hline
 \end{array} 
\cr \begin{array}{|cc|}
 \hline
 G_2\times SL_3(\RR) & SU(3)\times SL_3(\RR)\cr
 G_2\times SL_2(\RR) & SU(3)\times SL_2(\RR)\cr\hline
 \end{array}
 \end{align}
 It may also be helpful to display the generators for the four groups in a table
 \begin{align}\label{genstable}
 \begin{array}{|l|l|}
 \hline
 X_1,X_L,D_L & X_l,X_{lL},D_l\cr
 Y_1,Y_L,Z_1,Z_L,E_L & Y_l,Y_{lL},Z_l,Z_{lL},E_l\cr\hline
 X_K,X_{KL},D_K,D_{KL},D_{L,KL},D_{K,L},D_{K,KL} & X_{lK},X_{lKL}\cr
 Y_K,Y_{KL},Z_{K},Z_{KL},E_K,E_{KL} & Y_{lK},Y_{lKL},Z_{lK},Z_{lKL}\cr\hline
 \end{array}
 \end{align}
 where each box contains the extra generators needed in each case, to add to the generators of the group(s) above and/or to the left of the box.
 
 We would 
 expect to use $Spin(3,1)$ in the top right corner for the Lorentz group, acting on the four labels $u,U,l,L$ for spacetime and/or $4$-momentum.
 However, in the Dirac algebra including $\gamma_5$ we need $Spin(3,2)$, which appears in the bottom-left, acting on the five labels $u,U,K,L,KL$. From this labelling
  we see that we have lost one of the three dimensions of momentum, labelled $l$, and gained two dimensions of something else, labelled $K,KL$. In particular, we have
  broken the symmetry of spacetime, to include a preferred direction in space, and we have broken the symmetry of $I,J,K$, to obtain a weak force with
  a broken symmetry group $SL_2(\RR)$. The breaking of the $I,J,K$ symmetry seems most likely to be a breaking of the generation symmetry of
  fundamental fermions. 
  We can, of course, avoid the breaking of the spacetime-symmetry by extending to $Spin(4,2)$ and the complex Dirac algebra, as is effectively done in the Standard Model.

But  in the $E_8$ model it seems more likely that the symmetry-breaking from complex quaternions and $Spin(4,2)$
down to real split quaternions and $Spin(3,2)$
is important for the description of those parts of physics
that use (mathematically real) momentum but not (mathematically imaginary) current.
 In other words we may need this symmetry-breaking for the description of mass and gravity.
The particular space direction chosen must be something physically important such as the direction of acceleration relative to an inertial frame,
  or the direction of the ambient gravitational field, or the direction of the ambient angular momentum. 
   
This suggests that the top right corner, in the form
\begin{align}
SL_3(\CC) \times SU(3) \times SL_3(\RR),
\end{align}
 represents the Standard Model of elementary particles, with three colours of quarks and three generations of fermions (but only for half of the right-handed quarks!).
 The restriction from $SL_3(\CC)$ to $SL_2(\CC)$ splits fermions from bosons, and allows every observer to choose their own preferred copy of the
 Lorentz group $SL_2(\CC)$. The ten remaining dimensions of $SL_3(\CC)$ consist of a complex scalar and a Dirac spinor. Hence there is an $8$-parameter
 family of copies of $SL_2(\CC)$ available for different observers. This compares to a $9$-parameter family of copies of $SO(3,1)$ inside $SL_4(\RR)$ that
 describes the analogous phenomenon in General Relativity. Clearly, therefore, this model does not resolve the basic problem of incompatibility of GR with QM.
 It does, however, provide an explanation of sorts for the so-called `right-handed neutrinos': these are not interpreted as particles, but as transformations
 between different coordinate systems preferred by different observers.

The interpretation of the bottom left corner, in the form
\begin{align}
Sp_6(\RR) \times G_2 \times SL_2(\RR),
\end{align} 
is now freed from the necessity to include the Lorentz group in $Sp_6(\RR)$, and hence freed from the necessity to extend Minkowski spacetime, with symmetry group 
$SO(3,1)$, to
anti-de Sitter spacetime, with symmetry group $SO(3,2)$. Indeed, the latter group, or rather its double cover $Sp_4(\RR)$, now only has to act on two of the three
spatial coordinates, and can therefore plausibly be identified with the group of symmetries of phase space for $2$-dimensional dynamics. The choice of which two
dimensions these are is the same as the choice of restriction from $SO(3,1)$ to $SO(2,1)$ above, and therefore has the same relationship to
acceleration, rotation and/or the gravitational field. In practice, most experiments are horizontal, so that the most likely direction in most cases will be
the direction of the gravitational field. However, other directions may enter into the model at various points.
 
Finally, the extension from $Sp_4(\RR)$ to $Sp_6(\RR)$ allows individual observers to choose whichever $2$-dimensional part of dynamics they wish to model
with the Standard Model, and hence to modify the Standard Model for different ambient conditions of acceleration, rotation and gravity. 
If we want a model that is truly relativistic (independent of the observer), then we can treat $Sp_6(\RR)$ as the symmetry group of phase space for
$3$-dimensional dynamics, which entails abandoning the concept of `spacetime', since it is now observer-dependent, and therefore no longer useful for 
a fundamental theory. The embedding of $Sp_4(\RR)$ via $Sp_4(\RR) \times Sp_2(\RR)$ into $Sp_6(\RR)$ shows that there is again an
$8$-parameter family of copies of $Sp_4(\RR)$ available, for an $8$-parameter family of observers.
This extends to an $11$-parameter family of copies of $SL_2(\CC)$, in case this is useful.

A more radical proposal is made in \cite{CAMPS}, in which it is noted that both $SU(3)$ and $SL_3(\RR)$ are subgroups of $Sp_6(\RR)$, and it is
therefore proposed to identify the colour symmetry group $SU(3)$ and the generation symmetry group $SL_3(\RR)$ with the corresponding subgroups
of $Sp_6(\RR)$, so that the latter group now contains all of the required symmetries of fundamental physics. It may be that this proposal is too radical, but it is certainly
necessary to have \emph{some} mechanism for linking the generation symmetry group $SL_3(\RR)$ to \emph{some} concept of mass.

\section{Gravity}
\label{gravity}
It must be stressed again that with the standard interpretation of the Lorentz group $SL_2(\CC)$ acting on spacetime, labelled by $u,U,l,L$, there is no conceivable way
of implementing General Relativity inside $E_{8(-24)}$. The fact that $SL_4(\RR)$ has no double cover that acts on
spinors categorically rules this out. However, there are bits of $E_8$ that appear not to be used in the Standard Model, that could in principle be used for
a quantum theory of gravity, and that might approximate to GR in appropriate circumstances. Or it may be that a suitable tweak to the interpretation
might allow GR in to the model.

For example, $SL_4(\RR)$ acting on spacetime translates to $SO(3,3)$ acting on phase space, and there is an obvious copy of $SO(3,3)$ in $SU(3,3)$, that
extends $SL_3(\RR)$ in $Sp_6(\RR)$. This suggests that complexifying phase space, in order to implement momentum and current separately, may be key
to including quantum gravity in the model. In other words, $A_5$ is the battleground in which the QM and GR models must resolve their differences.
These differences mainly lie in the fundamental properties of spacetime, so the hope is that by shifting the emphasis from spacetime to phase space,
a compromise might be reached in which spacetime is never defined or used at all.
If QM uses $Sp_6(\RR)$ to act on phase space, as it should in a Hamiltonian theory, and GR uses $SO(3,3)$, then both can be embedded in $SU(3,3)$,
and each can be allowed to add corrections to the other. 

The key to this process is to re-interpret the various spin groups as acting on $2+2$-dimensional (real, complex or quaternionic) phase space, and not on an abstract
space of `spinors' that have no concrete physical interpretation. We have already done this \cite{CAMPS}
with the group $Spin(3,2)\cong Sp_4(\RR)$ in $C_3$, interpreted as
a 
subalgebra 
of the Dirac algebra, so we must now do the same with $Spin(4,2)\cong SU(2,2)$ in $A_5$, and interpret this both as
the full complex Dirac algebra, and as a complex version of the subgroup $SO(2,2)$ of $SO(3,3)$, corresponding to a subgroup $SL_2(\RR) \times SL_2(\RR)$ of $SL_4(\RR)$
in GR.

Within the $2\times 2$ magic square, we therefore have a copy of 
\begin{align}
SO(2,2)=Spin(2,1)\otimes Spin(2,1)
\end{align} 
inside $Spin(4,2)$, from which
we can write down generators for $SO(3,3)$ as follows:
\begin{align}
\label{D3gens}
D_L, D_{K,KL}, X_1,X_L,X_{lK},X_{lKL},\cr
E_L,Y_1,Y_L,Y_{lK},Y_{lKL},\cr
Z_1,Z_L,Z_{lK},Z_{lKL}
\end{align}
These are the elements in the top left and bottom right of (\ref{genstable}), together with $D_{K,KL}$.
The first row of (\ref{D3gens}) consists of generators for $SO(2,2)$. The elements whose labels do not include $l$ generate $GL_3(\RR)$ inside $Sp_6(\RR)$.

This choice of $SO(2,2)$ amounts to splitting the $6$ labels into $u,U,L$ for one copy of $Spin(2,1)$ and $l,K,KL$ for the other. 
In matrix terms we have
\begin{align}
\label{D2mats}
&\begin{pmatrix}L&\cr 0&-L\end{pmatrix},
\begin{pmatrix}0&1\cr -1&0\end{pmatrix},
\begin{pmatrix}0&L\cr L&0\end{pmatrix},\cr
&\begin{pmatrix}L&0\cr 0&L\end{pmatrix},
\begin{pmatrix}0&lK\cr -lK&0\end{pmatrix},
\begin{pmatrix}0&lKL\cr -lKL&0\end{pmatrix},
\end{align}
where the first row is $SO(2,1)$ acting on $2+1$-dimensional spacetime.

This splitting is similar to Woit's splitting \cite{WoitSO24,WoitRH} into `right-handed' spacetime and `left-handed' gauge groups, 
respectively, except that he complexifies everything and interprets these
groups as compact $SU(2)$ rather than split $SL_2(\RR)$. 
Moreover, he interprets $SU(2)$ as acting on a Euclidean rather than Lorentzian spacetime, or a spinor.
In our interpretation, the `right-handed' copy acts on $2+1$-spacetime, which we must presumably interpret as being
perpendicular to the local direction of the gravitational field, and the `left-handed' copy acts on $2+1$ fermions in a single generation.

At this point we may notice that we have acquired two copies of $SL_2(\RR)$, both labelled `left-handed', one acting on labels $l,K,KL$, the other acting
on labels $I,J,IL,JL$. Some mixing of the two may be required in order to define masses for the intermediate vector bosons.
We also have a second `right-handed' copy acting on the six labels $I,J,K,IL,JL,KL$, but not acting on $l$.
Again it is not clear whether there is any relationship between this right-handed copy, analogous to the one defined in \cite{MDW}, and the
one acting on $u,U,L$ (with or without $l$), analogous to the one defined in \cite{WoitSO24,WoitRH}.
What is clear, at least, is that the terms `right-handed' and `left-handed' have multiple different meanings in the literature,
and are consequently best avoided.

\section{General relativity}
\label{GR}
The tensors used in GR are irreducible representations of $SO(3,3)$ of dimensions $6$ (field strength tensor), $10$ (Ricci tensor, stress-energy tensor) and $20$
(Riemann curvature tensor). The representations of $SU(3,3)$ that are available in $E_8$ are complex $\rep6_\CC$ and $\rep{15}_\CC$, and real $\rep{20}$ 
and $\rep{35}$, which restrict
to $SO(3,3)$ as follows:
\begin{align}
\rep 6_\CC & \rightarrow \rep 6\cr
\rep {15}_\CC & \rightarrow \rep{15}\cr
\rep{20} & \rightarrow\rep{10}a+\rep{10}b \cr
\rep{35} & \rightarrow \rep{15}+\rep{20}
\end{align}
It follows that the Einstein field equations must be written in terms of the representation $\rep2\otimes \rep{20}$ on left-handed leptons, that relates matter to neutrinos,
and therefore expresses the gravitational field in terms of neutrinos. On the other hand, the Riemann Curvature Tensor (RCT)  lies in the adjoint representation
of $SU(3,3)$, and consists of the $20$ dimensions outside $SO(3,3)$. Thus the RCT becomes part of the symmetry group of the model,
and consists of the elements in the top right and bottom left of (\ref{genstable}), excluding the `scalar' $D_{K,KL}$. Finally, the field strength tensor 
is coupled to the quark colour/charge representation $\rep2\otimes \rep3_\CC$, and hence to baryonic matter. 

From this analysis, we see that the only essential thing that is missing from GR is the distinction between $\rep{10}a$ and $\rep{10}b$, which are usually regarded as being
self-dual and therefore equivalent, but are in fact dual \emph{to each other}. 
In the usual formalism in terms of the Lorentz group $SO(3,1)$, they both restrict to $\rep1+\rep9$, so that the distinction between them is less obvious.
The introduction of a second scalar (the cosmological constant) only extends from $10$ to $11$ variables, when $20$ are required for the full theory.
A consequence of this generalisation is that electrons are far more important for
gravity than is usually supposed, since all three generations of neutrinos and antineutrinos, and therefore all three generations of electrons, participate in an essential way.

Indeed, it is possible to include the whole of GR and the weak force inside $E_6$ by adding the representation $\rep2\otimes\rep{20}$ to the adjoint
representation of $SU(3,3)$, so that on restriction to
$A_1+A_3$ the adjoint representation of $E_6$ has the structure
\begin{align}
\rep 3\otimes \rep1 + \rep1\otimes (\rep{15} + \rep{20}) + \rep2\otimes (\rep{10}a+\rep{10}b).
\end{align}
The representation $\rep3\otimes\rep1$ is the adjoint representation of $SL_2(\RR)$, generated by
\begin{align}
D_{I,J}- D_{IL,JL},\quad  D_{I,IL}- D_{J,JL},\quad  D_{J,IL}+ D_{I,JL}.
\end{align}
and generators for the adjoint representation $\rep1\otimes \rep{15}$ of $SO(3,3)$ are listed in (\ref{D3gens}).

A translation to 
$6\times 6$ real matrices is obtained via the map
\begin{align}
lKL\mapsto \begin{pmatrix}0&1\cr -1&0\end{pmatrix},\quad L\mapsto \begin{pmatrix}0&1\cr 1&0\end{pmatrix},\quad lK\mapsto \begin{pmatrix} 1 & 0\cr 0&-1\end{pmatrix}
\end{align}
but it is noteworthy that this translation does \emph{not} map $\rep{15}$ to antisymmetric matrices, 
as one might naively expect. This is because of the difference between the split and compact signatures.
For example, we have
\begin{align}
&X_1\mapsto \begin{pmatrix} 0&0&1&0\cr0&0&0&1\cr -1&0&0&0\cr 0&-1&0&0\end{pmatrix},
X_L\mapsto \begin{pmatrix} 0&0&0&1\cr 0&0&1&0\cr 0&1&0&0\cr 1&0&0&0\end{pmatrix},\cr
&X_{lK}\mapsto \begin{pmatrix}0&0&1&0\cr 0&0&0&-1\cr -1&0&0&0\cr 0&1&0&0\end{pmatrix},
X_{lKL}\mapsto \begin{pmatrix}0&0&0&1\cr 0&0&-1&0\cr 0&-1&0&0\cr 1&0&0&0\end{pmatrix}.
\end{align}

The Riemann Curvature Tensor $\rep{20}$ is generated by 
\begin{align}
\label{RCTgens}
&D_l,E_l,D_K,E_K,D_{KL},E_{KL},D_{L,KL},D_{K,L},\cr
 &X_l,X_{lL},X_{K},X_{KL},\cr
&Y_l,Y_{lL},Y_{K},Y_{KL},\cr
&Z_l,Z_{lL},Z_{K},Z_{KL}.
\end{align}
We can take the five elements with label $l$ to correspond to the diagonal elements of the tensor as $6\times 6$ matrices.

The four $10$-dimensional tensors consist of all the elements containing $I$ or $J$ in their labels, and they each split into $D$s, $X$s, $Y$s and $Z$s as $4+2+2+2$.
Bases for the four representations are given below. The top two can be taken as the two copies of $\rep{10}a$, mixed by $SL_2(\RR)$, in which case the bottom two are
$\rep{10}b$.
\begin{align}
&\begin{array}{|c|}
\hline
X_{lI(1+L)}, Y_{lI(1+L)}, Z_{lI(1+L)},\cr
X_{J(1+L)},Y_{J(1+L)},Z_{J(1+L)},\cr
D_{K,I(1+L)},E_{K,I(1+L)},\cr
D_{KL,I(1+L)},E_{KL,I(1+L)}.\cr\hline
\end{array}\quad
\begin{array}{|c|}
\hline
X_{lI(1-L)}, Y_{lI(1-L)}, Z_{lI(1-L)},\cr
X_{J(1-L)},Y_{J(1-L)},Z_{J(1-L)},\cr
D_{K,I(1-L)},E_{K,I(1-L)},\cr
D_{KL,I(1-L)},E_{KL,I(1-L)}.\cr\hline
\end{array}\cr
&\begin{array}{|c|}
\hline
X_{I(1+L)}, Y_{I(1+L)}, Z_{I(1+L)},\cr
X_{lJ(1+L)},Y_{lJ(1+L)},Z_{lJ(1+L)},\cr
D_{K,J(1+L)},E_{K,J(1+L)},\cr
D_{KL,J(1+L)},E_{KL,J(1+L)}.\cr\hline
\end{array}\quad
\begin{array}{|c|}
\hline
X_{I(1-L)}, Y_{I(1-L)}, Z_{I(1-L)},\cr
X_{lJ(1-L)},Y_{lJ(1-L)},Z_{lJ(1-L)},\cr
D_{K,J(1-L)},E_{K,J(1-L)},\cr
D_{KL,J(1-L)},E_{KL,J(1-L)}.\cr\hline
\end{array}
\end{align}

\section{MOND}
The proposed extension to General Relativity involves extending the standard $10$-dimensional symmetric rank $2$ tensors on spacetime to $20$-dimensional
anti-symmetric rank $3$ tensors on phase space. Restricting to the Lorentz group $SO_3(\CC)$, the latter breaks up as $1+9+9+1$, containing
two real scalars, with spin $(0,0)$, and two real representations with spin $(1,1)$. With respect to a basis of phase space consisting of $3$ directions
of momentum and $3$ directions of position, the antisymmetric cube breaks up into pieces with $0$, $1$, $2$ and $3$ position coordinates,
of dimensions $1$, $9$, $9$ and $1$ respectively.

Newton--Einstein gravity consists mainly of the inverse-square law, with two position vectors, and a relativistic `mass' (or energy) that involves 
the momentum vector. This occupies one of the $9$-dimensional pieces, mixed with a scalar `rest mass', that involves all three position coordinates.
In other words, the rest mass is defined by an inverse cube law that relates an object to its surroundings. The other scalar has no position coordinates at all,
and therefore plays the role of a cosmological constant, or a `dark energy' field that in principle may not be constant throughout the universe
\cite{nodarkenergy}. Indeed, this
`dark energy' depends on the anti-symmetric cube of momentum, and therefore relates an object to the rest of the universe via Mach's Principle.

This leaves one $9$-dimensional piece of the representation unaccounted for in Newton--Einstein gravity. This has a single position vector, and therefore 
corresponds to an inverse-linear term in the universal law of gravity. Such a term was first proposed by Milgrom \cite{Milgrom1} 
in 1983, in order to reconcile the theory of gravity
with observations that appeared to be inconsistent with the 
existing theory.
This proposed modification to Newtonian dynamics (MOND) has since been
confirmed in a wide variety of studies of astronomical objects on varying scales \cite{MOND,Chae,Hernandez}.

The model proposed here suggests a dependence of this inverse-linear term on a square of momentum, which 
may mean the momentum of each body 
relative to the `centre-of-mass' frame, assuming that such a frame can be adequately defined. 
However, it may be necessary to embed the system in a larger system (or external field), 
 in which case
we have two independent momenta relative to the external system.
In particular, for a star on the outskirts of a spiral galaxy, with a 
reasonably `standard' stellar mass, and where the speed of rotation with respect to the centre of the galaxy is observationally close to a constant value (in contrast to the
expected Keplerian decline), this momentum-squared factor is close to being a constant. But the point at which the inverse-linear term becomes
dominant is not yet clear from this picture, and I have not found a convincing explanation for the empirical `critical acceleration' $a_0$.
What appears to be critical is a ratio of momentum to distance, which translates in Newtonian physics to a ratio of mass to duration.

A number of models \cite{Sanders,TeVeS,AeST}
have attempted to reconcile GR with MOND by adding extra scalar and/or vector fields to the Einsteinian tensors. However, the analysis of
symmetries carried out in this paper suggests that it is not sufficient to add scalars or vectors, but that it is necessary to add a second tensor,
dual to the first. On the other hand it has been proposed by Yahalom \cite{Yahalom} that the inverse-linear term 
is merely an artefact of the 
fact that the gravitational field propagates 
at a finite and not infinite speed. In other words, it is not enough to know where an object is, it is also necessary to know how fast its mass is travelling.
It may or may not be necessary to separate its momentum into mass and velocity, but it is certainly necessary 
to know the momentum. Thus it may be possible to implement this `retarded gravity' approach within the phase space model
that I propose.

\section{Conclusion}
The problem of unification of particle physics and gravity goes back almost a century, and occupied Einstein for at least a quarter of that century. Yet the problem
seems no nearer to a solution today than it did fifty years ago, and even further away than it seemed forty or thirty years ago. This indicates that there must be
something subtly wrong in the basic assumptions somewhere. My analysis locates this problem in the concept of spacetime itself. The way that spacetime
is treated in relativity, using the Lorentz group in the form $SO(3,1)$, is \emph{mathematically} (never mind physically)
inconsistent with the way that spacetime is treated in quantum mechanics, using
the Lorentz group in the form $SL_2(\CC)$. It isn't a question of one of them being `right' and the other one `wrong', it is a question of there being no
consistent definition of spacetime at all, and no possible way to measure spacetime in the absence of objects embedded in spacetime.

Therefore I have considered the possibility of describing physics without using spacetime, but instead using only phase space, as Hamilton taught us to do.
This involves re-interpreting the Dirac algebra, the Einstein field equations and the Riemann curvature tensor in terms of a complex phase space, in order to
include both momentum and current. 
The symmetry group of the complex phase space in three dimensions is the complex symplectic group $Sp_6(\CC)\cong SU(3,3)$.
Taking my cue from the variety of models based on $E_8$ and the Freudenthal--Tits magic square, from $E_8\times E_8$ heterotic 
string theory down, I focus on the group $SU(3,3)$ embedded in $E_{8(-24)}$ via $E_{6(2)}$, and find within these groups 
all the mathematical structures that a unified model appears to require.

This is not, of course, in itself a unified theory of fundamental physics.
But it is a unified mathematical model, in which all the ingredients of all the fundamental theories of physics can be found.
This includes the complex Dirac algebra, the gauge groups of the weak and strong nuclear forces, including symmetry-breaking of the weak force,
a classification of elementary fermions in which there are no right-handed neutrinos, general covariance, and all the tensors used in GR.
Moreover, these ingredients fit together in ways that are broadly consistent with experiment.
I therefore suggest that this is a promising foundation on which to try to build a unified theory of fundamental physics.


\begin{thebibliography}{9(}
\bibitem{MDW} C. A. Manogue, T. Dray and R. A. Wilson (2022), Octions: an $E_8$ description of the standard model,
{\it J. Math. Phys.} {\bf 63}, 081703.
arXiv:2204.05310


\bibitem{Lisi} 
 A. G. Lisi (2007), An exceptionally simple theory of everything,
 arXiv:0711.0770
\bibitem{Lisi2}
A. G. Lisi (2010), An explicit embedding of gravity and the standard model in E8,
arXiv:1006.4908

\bibitem{Chester} D. Chester, A. Marrani and M. Rios (2020),
Beyond the standard model with six-dimensional spacetime,
arXiv:2002.02391.

\bibitem{chirality} R. A. Wilson (2022), Chirality in an $E_8$ model of elementary particles,
arXiv:2210.06029.



\bibitem{DG} J. Distler and S. Garibaldi (2010), There is no $E_8$ theory of everything,
{\it Communications in Math. Phys.} {\bf 298} (2), 419--436.

\bibitem{Freud} H. Freudenthal, Beziehungen der $E_7$ und $E_8$ zur Oktavenebene. X-XI. 
{\it Indag. Math.} {\bf 25} (1963), 457--487.

\bibitem{Tits} J. Tits, Alg\`ebres alternatives, alg\`ebres de Jordan et alg\`ebres de Lie  exceptionelles,
{\it Indag. Math.} {\bf 28} (1966), 223--237.

\bibitem{BS} C. H. Barton and A. Sudbery,
Magic squares and matrix models of Lie algebras,
{\it Advances in Math.} {\bf 180} (2003), 596--647.


\bibitem{MD} C. A. Manogue and T. Dray (2010), 
Octonions, E6, and particle physics, {\it  J. Phys. Conf. Ser.} {\bf 254}, 012005.

\bibitem{Todorov} I. Todorov and M. Dubois-Violette (2018), 
Deducing the symmetry of the standard model from the automorphism and structure groups of the exceptional Jordan algebra,
{\it Int. J. Mod. Phys. A} {\bf 33}, 1850118
  
\bibitem{PatiSalam} J. C. Pati and A. Salam (1974), Lepton number as the fourth `color',
{\it Phys. Rev. D} {\bf 10} (1), 275--289.

\bibitem{twistors} R. Penrose (1967), Twistor algebra, {\it J. Math. Phys.} {\bf 8} (2), 345--366.


\bibitem{twistorlectures} T. Adamo (2017), {\it Lectures on twistor theory}.
arXiv:1712.02196.

\bibitem{Maldecena} J. Maldecena (1998), The large $N$ limit of superconformal field theories and supergravity,
{\it Adv. Theor. Math. Phys.} {\bf 2}, 231--252.

 \bibitem{DMWE7a} T. Dray. C. A. Manogue and R. A. Wilson (2014),
 A symplectic representation of $E_7$,
 {\it Comment. Math. Univ. Carolin.} {\bf 55}, 387.
  
 \bibitem{DMWE7} T. Dray, C. A. Manogue and R. A. Wilson (2024),
 A new division algebra representation of $E_7$,
 arXiv:2401.105

\bibitem{deGosson} M. de Gosson and B. Hiley (2011), Imprints of the quantum world in classical mechanics,
{\it Found. Phys.} {\bf 41}, 1415--1436.

\bibitem{WDM} 
 R. A. Wilson, T. Dray and C. A. Manogue (2022), An octonionic construction of $E_8$ and the Lie algebra
 magic square, 
arXiv:2204.04996


\bibitem{GellMann} M. Gell-Mann (1961), The eightfold way: a theory of strong interaction symmetry,
Synchrotron Lab. Report CTSL-20, Cal. Tech.


\bibitem{QCD} W. Greiner, S. Schramm and E. Stein (2007), {\it Quantum Chromodynamics}, Springer.


\bibitem{Coleman} S. Coleman and J. Mandula (1967), All possible symmetries of the S matrix,
{\it Physical Review} {\bf 159} (5), 1251.


\bibitem{WoitSO24} P. Woit (2021), Euclidean spinors and twistor unification, arXiv:2104.05099

\bibitem{WoitRH} P. Woit (2023), Spacetime is right-handed, arXiv:2311.00608.

\bibitem{CAMPS} R. A. Wilson (2024),
A Clifford algebra model in phase space,
arXiv:2404.04278.

\bibitem{nodarkenergy} J. Colin, R. Mohayaee, M. Rameez and S. Sarkar (2019),
Evidence for anisotropy of cosmic acceleration,
{\it Astronomy and Astrophysics} {\bf 631}, L13.
arXiv:1808.04597.

\bibitem{Milgrom1}  M. Milgrom (1983),
A modification of the Newtonian dynamics as a possible alternative to the
hidden mass hypothesis,
{\it Astrophysical J.} {\bf 270},  365--370.

\bibitem{MOND} B. Famaey and S. McGaugh (2012), Modified Newtonian dynamics (MOND):
observational phenomenology and relativistic extensions, {\it Living reviews in relativity} {\bf 15}, 10.
arXiv:1112.3960.


\bibitem{Chae} K.-H. Chae (2023), Breakdown of the Newton--Einstein standard gravity at low accelerations in 
internal dynamics of wide binary stars,
{\it The Astrophysical Journal} {\bf 952}, 128.

\bibitem{Hernandez} X. Hernandez (2023), {\it MNRAS} {\bf 525}, 1401.


\bibitem{Sanders} R. H. Sanders (1997), A stratified framework for scalar-tensor theories of Modified Dynamics,
{\it Astrophys. J.} {\bf 480}, 492--502.

\bibitem{TeVeS} J. D. Bekenstein (2004), Relativistic gravitation theory for the modified Newtonian dynamics paradigm,
{\it Phys. Rev. D} {\bf 70}, 083509.

\bibitem{AeST} C. Skordis and T. Z\l o\' snik (2021), A new relativistic theory for
Modified Newtonian Dynamics,
{\it Phys. Rev. Lett.} {\bf 127}, 161302.


\bibitem{Yahalom} A. Yahalom (2019), The effect of retardation on galactic rotation curves,
{\it J. Phys.: Conf. Ser.} {\bf 1239}, 012006.



\end{thebibliography}
\end{document}